\documentclass[aps,prl,nofootinbib,twocolumn,superscriptaddress,preprintnumbers,floatfix]{revtex4-2} 
\usepackage[nodisplayskipstretch]{setspace}
\usepackage{mathtools}
\usepackage{xcolor}
\usepackage{amssymb}
\usepackage{amsmath}
\usepackage{hyperref}
\usepackage{breakurl}
\usepackage[english]{babel}
\begin{document}

\title{\sffamily\bfseries Conservative binary dynamics to third post-Minkowskian order beyond General Relativity}

\author{Gabriel Luz Almeida}
\affiliation{School of Physics \& Shing-Tung Yau Center, Southeast University, Nanjing 210096, China}

\author{Yuchen Du}
\affiliation{School of Physics \& Shing-Tung Yau Center, Southeast University, Nanjing 210096, China}

\author{Zhengwen Liu}
\affiliation{School of Physics \& Shing-Tung Yau Center, Southeast University, Nanjing 210096, China}
\affiliation{Niels Bohr International Academy \& Center of Gravity, Niels Bohr Institute, University of Copenhagen, Blegdamsvej 17, 2100 Copenhagen, Denmark}

\author{Hongbin Wang}
\affiliation{School of Physics \& Shing-Tung Yau Center, Southeast University, Nanjing 210096, China}

\begin{abstract}
We present the conservative dynamics of compact binaries to third order in the post-Minkowskian approximation in a theory that extends general relativity by a massless scalar field coupled to the Gauss-Bonnet invariant. We employ the effective field theory approach to construct the effective action of binary systems by integrating out the metric and scalar degrees of freedom that mediate the gravitational interactions between the two bodies. We derive analytical expressions for the scattering impulse and the deflection angle to third order in the post-Minkowskian expansion. Our results are found to be in agreement, in the overlapping regimes, with state-of-the-art calculations in the post-Newtonian/post-Minkowskian theory.
\end{abstract}
\maketitle

\noindent{\bf Introduction.}~The direct detection of gravitational waves provides a unique opportunity to test gravity with unprecedented precision and to probe potential deviations signaling new physics beyond Einstein's theory of general relativity \cite{LIGOScientific:2016lio,LIGOScientific:2018dkp,LIGOScientific:2021sio,LIGOScientific:2025wao}.
Meeting this challenge requires that the theoretical predictions for the dynamics of compact binary systems match the accuracy of current and, in particular, forthcoming gravitational-wave observatories, including LIGO-Virgo-KAGRA, the Einstein Telescope, and LISA.
This demand has stimulated extensive efforts to develop a variety of complementary approaches to the relativistic two-body problem, including numerical relativity \cite{Pretorius:2005gq,Campanelli:2005dd,Baker:2005vv,Damour:2014afa,Boyle:2019kee}, the post-Newtonian (PN) \cite{Einstein:1938yz,Ohta:1973je,Blanchet:2013haa,Damour:2019lcq} and post-Minkowskian (PM) \cite{Bertotti:1956pxu,Kerr:1959zlt,Bertotti:1960wuq,Westpfahl:1979gu,Portilla:1980uz,Bel:1981be} approximations, gravitational self-force formalism \cite{Mino:1996nk,Quinn:1996am,Poisson:2011nh,Barack:2018yvs,Gralla:2021qaf}, the effective-one-body framework \cite{Buonanno:1998gg,Buonanno:2000ef}, and the boundary-to-bound correspondence \cite{Kalin:2019rwq,Kalin:2019inp,Cho:2021arx}.
Recently, the quantum field theory (QFT) methodology (including effective field theory (EFT) \cite{Goldberger:2004jt,Porto:2016pyg,Kalin:2020mvi,Kalin:2020fhe,Dlapa:2023hsl} and scattering amplitude \cite{Bjerrum-Bohr:2018xdl,Cheung:2018wkq,Bern:2019nnu,Bern:2019crd,DiVecchia:2023frv,Kosower:2018adc} methods), powered by modern Feynman loop integration technology \cite{Henn:2013pwa,Lee:2014ioa,Dlapa:2020cwj,Dlapa:2022wdu,Jinno:2022sbr,Smirnov:2023yhb,Lee:2013mka,Bree:2025tug,Weinzierl:2022eaz,Duhr:2019tlz}, has proven particularly powerful in solving the two-body problem within both the PM and PN expansions.
In general relativity, these methods have enabled the high-order computation of the two-body dynamics, now known up to fifth PM order \cite{Dlapa:2025biy,Dlapa:2023hsl,Dlapa:2024cje,Dlapa:2022lmu,Dlapa:2021vgp,Dlapa:2021npj,Liu:2021zxr,Kalin:2020lmz,Kalin:2020fhe,Cheung:2018wkq,Driesse:2026qiz,
Driesse:2024feo,Driesse:2024xad,Jakobsen:2023hig,Jakobsen:2023ndj,Jakobsen:2022psy,Bern:2025wyd,Bern:2025zno,Bern:2024adl,Bern:2022kto,Bern:2021yeh,Bern:2021dqo,Bern:2020buy,Bern:2019crd,Bern:2019nnu,Cheung:2020sdj,Cheung:2020gyp,Bern:2019crd,Bjerrum-Bohr:2018xdl,Cristofoli:2019neg,Bjerrum-Bohr:2019kec,Bjerrum-Bohr:2021vuf,Bjerrum-Bohr:2021din,Bjerrum-Bohr:2021wwt,Damgaard:2023ttc,Damgaard:2023vnx,Damgaard:2021ipf,Heissenberg:2025ocy,DiVecchia:2023frv,DiVecchia:2022piu,DiVecchia:2021bdo,DiVecchia:2021ndb}, as well as to fourth PN order (4PN) \cite{Damour:2014jta,Galley:2015kus,Marchand:2017pir,Foffa:2019rdf,Foffa:2019yfl,Cho:2022syn,Blanchet:2023sbv} (with partial results also available in higher orders \cite{Brunello:2025gpf,Foffa:2019hrb,Blumlein:2019zku,Almeida:2021xwn,Blumlein:2021txj,Porto:2024cwd}).

\vskip 4pt
While precision predictions within general relativity remain foundational for testing gravitational theories with gravitational waves, waveform modeling in alternative theories is equally crucial, particularly for theory-specific tests, which provide essential complementarity to theory-independent analyses.
Within this context, scalar-tensor theories and Einstein-scalar-Gauss-Bonnet (ESGB) gravity stand out as particularly well-motivated and phenomenologically rich candidates.
The scalar-tensor theories constitute some of the simplest and best-studied extensions of general relativity.
ESGB theory, in turn, can be viewed as a curvature-coupled realization within this broader scalar-tensor sector, belonging to the Horndeski class and modifying compact-object dynamics through a scalar coupling to the Gauss-Bonnet invariant.
In particular, this additional interaction allows black holes to develop nontrivial scalar configurations \cite{Sotiriou:2014pfa,Sotiriou:2013qea}, in contrast to standard massless scalar-tensor theories where stationary black holes coincide with their general-relativistic counterparts \cite{Hawking:1972qk}.
This distinction is especially relevant given that most observed compact-binary mergers involve black-hole systems.
These theories provide a controlled framework to investigate strong-field deviations from general relativity and their imprints on gravitational-wave signals.

\vskip 4pt
Motivated by these features, mergers of black hole and neutron star binaries in scalar-tensor and ESGB gravity have been extensively investigated \cite{Wagoner:1970vr,
Will:1994fb,Berti:2015itd,Blazquez-Salcedo:2016enn,Langlois:2017dyl,Baker:2017hug,Witek:2018dmd,Berti:2018cxi,Lyu:2022gdr}. Nevertheless, unlike in general relativity, analytic studies of the relativistic two-body problem in these theories remain largely restricted to the post-Newtonian regime. A useful snapshot of the current analytic status is provided by the construction of the first complete inspiral-merger-ringdown waveform model in ESGB gravity \cite{Julie:2024fwy}, built upon the conservative dynamics through 3PN order and the associated effective-one-body Hamiltonian \cite{Damour:1992we,Mirshekari:2013vb,Almeida:2024uph,Bernard:2018hta,Bernard:2018ivi,Julie:2022qux}, metric waveform amplitudes for nonspinning binaries through 2PN  \cite{Lang:2013fna,Lang:2014osa} and at 2.5PN for scalar waves beyond leading order in scalar-tensor theories \cite{Bernard:2022noq}. Further analytic progress in ESGB gravity includes EFT derivations of spin-dependent corrections to the conservative dynamics through 3PN order \cite{Almeida:2024cqz} and radiative results beyond the quadrupole at first PN accuracy \cite{Shiralilou:2021mfl}.

\vskip 4pt
As emphasized, modern QFT techniques have demonstrated remarkable promise in addressing the two-body problem, yielding a wealth of high-order results in general relativity.
This success naturally motivates extending such methods beyond Einstein gravity.
In this work, we take a step in that direction: focusing on massless scalar-tensor and ESGB gravity, and employing the EFT approach, we compute the conservative momentum impulse and scattering angle for compact binary systems through third order in the PM expansion.

\vskip 6pt
\noindent{\bf Framework.}
We consider the theory with the action
\begin{align}\label{action-ESGB}
S_\text{g} =  - {2\over \kappa^2}  \int \mathrm{d}^4 x \sqrt{-g}
\Big(
R - 2(\partial\varphi)^2
- \alpha f(\varphi) \mathcal{G}
 \Big),
\end{align}
where $R$ is the Ricci scalar, $\varphi$ is a massless real scalar, and $\kappa = M_\textrm{Pl}^{-1} = \sqrt{32\pi G}$ with $G$ Newton's constant.
We use $\eta_{\mu\nu} =\operatorname{diag}(1,-1,-1,-1)$ for the Minkowski metric.
The Gauss-Bonnet invariant, defined as $\mathcal{G} = R^2 - 4R_{\mu\nu}R^{\mu\nu} + R_{\mu\nu\rho\sigma}R^{\mu\nu\rho\sigma}$, which reduces to a topological surface term in four dimensions, is coupled to $\varphi$ through a function $f(\varphi)$, which specifies the model. 
A variety of coupling functions $f(\varphi)$, motivated by phenomenological applications and theoretical considerations, have been extensively studied in the literature \cite{Silva:2018qhn,Collodel:2019kkx,East:2021bqk,Doneva:2017bvd,Silva:2017uqg,Doneva:2022ewd}.
For instance, the choice $f(\varphi)=e^{\varphi}$ gives rise to Einstein-dilaton-Gauss-Bonnet gravity, which emerges as a low-energy effective description in heterotic string theories \cite{Gross:1986mw,Metsaev:1987zx}, while certain $f(\varphi)$ functions can lead to interesting phenomena such as spontaneous scalarization of black holes \cite{Doneva:2017bvd,Silva:2017uqg}.
In this work, we do not restrict to any particular functional form and keep $f(\varphi)$ general.

\vskip 5pt
We describe the self-gravitating compact objects using the point-particle approximation, where in the presence of scalar field couplings, such objects can be \textit{skeletonized} as scalar-field-dependent point masses $M(\varphi)$ \cite{Eardley:1975fgi,Damour:1992we}.
The resulting point-particle action takes the form
\begin{align}\label{action-pp}
  S_\text{pp} = -{1 \over 2} \sum_{i=1,2}\int \mathrm{d}\tau_i\, M_i(\varphi)\bigg(g_{\mu\nu} {\mathrm{d} {x}_i^\mu \over \mathrm{d}\tau_i} {\mathrm{d} {x}_i^\nu \over \mathrm{d}\tau_i}  + 1\bigg),
\end{align}
with $\tau_i$ the proper time along the worldline of particle $i$.

\vskip 5pt
Here, we are interested in the regime of weak gravitational (metric and scalar) fields, allowing for a perturbative expansion around their background values.
The spacetime metric is expanded around the flat Minkowski metric as $g_{\mu\nu} = \eta_{\mu\nu} +  \kappa h_{\mu\nu}$, while the scalar field can be similarly decomposed as $\varphi = \varphi_0 + \kappa\phi$ \cite{Will:1993book}.
Through its dependence on $\varphi$, the particle mass admits an expansion,
\begin{align}\label{mass-expand}
M_i(\varphi)  = m_i\Big[1 &+ s_i (\kappa \phi) + {s_i^2 + s_i' \over 2} (\kappa \phi)^2 
\nonumber\\
&+ {s_i^3 + 3s_is_i' + s_i'' \over 6} (\kappa \phi)^3 + \cdots \Big],
\end{align}
where $m_i=M_i(\varphi_0)$
and the coefficients
\begin{align}\label{def-mass-sentivity}
s_i \!=\! {\mathrm{d}\ln M_i \over \mathrm{d}\varphi}\Big|_{\varphi_0},
~
s_i' \!=\! {\mathrm{d}^2\ln M_i \over \mathrm{d}\varphi^2}\Big|_{\varphi_0},
~
s_i'' \!=\!  {\mathrm{d}^3\ln M_i \over \mathrm{d}\varphi^3}\Big|_{\varphi_0}
\end{align}
define the scalar \textit{sensitivities} of the compact object and encode its response to the scalar field.
Similarly, the coupling function $f(\varphi)$ admits an expansion of the form
\begin{equation}\label{}
f(\varphi) =  f_1\kappa \phi + {f_2 \over 2} (\kappa\phi)^2 + {f_3 \over 6} (\kappa\phi)^3 
 + \cdots,
\end{equation}
with coefficients $f_i$ determined by the specific theory.

\begin{figure}[t]
\centering
\includegraphics[width=80mm]{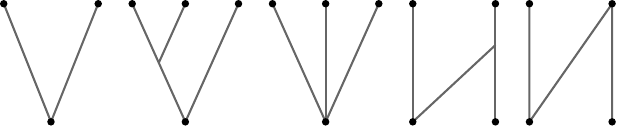}
\vskip -3mm
\caption{New Feynman topologies to 3PM.
Here black dots represent classical point-particle sources, and lines denote massless $h_{\mu\nu}$ or $\phi$ propagators.
In addition to the topologies arising in general relativity \cite{Kalin:2020fhe}, these new topologies appear due to the additional couplings between the scalar field $\phi$ and the point sources in scalar-tensor theory.}
\label{3PM-topo-new}
\vskip -12pt
\end{figure}

\vskip 4pt
The {\it classical effective action} can be obtained by \textit{integrating out} the propagating metric and scalar degrees of freedom in the path integral \cite{Goldberger:2004jt,Porto:2016pyg,Kalin:2020mvi,Dlapa:2023hsl}
\begin{align}\label{EFT}
e^{i S_\text{eff}[x_i]} = \int\mathcal{D}h\mathcal{D}\phi\, e^{ i S_\text{pp}[x_i,h,\phi] + iS_\text{g}[h,\phi] + iS_\text{gf}[h] },
\end{align}
in the saddle-point approximation, thereby eliminating quantum corrections.
The last term in the exponential on the right-hand side of the equation is a gauge-fixing term introduced to remove the diffeomorphism redundancy of the theory; physical observables are independent of its specific choice.
In practice, this integration is performed perturbatively in the post-Minkowskian expansion using Feynman diagrams constructed from the worldline couplings,  gravitational propagators ($h_{\mu\nu}$ and $\phi$), and interaction vertices derived from the expansion of the action \eqref{action-ESGB}.
Extremizing the resulting effective action yields the classical equations of motion for the two-body trajectories.
These equations are solved iteratively, starting with the unperturbed straight-line trajectories, $x_i^\mu = b_i^\mu + u_i^\mu\tau_i$, where $b_i^\mu$ and $u_i^\mu$ are the initial position and initial velocity of particle $i$, with $u_1^2 = u_2^2 = 1$.
From the evolved trajectories we can extract the scattering observables.
A key quantity is the impulse, defined as the total change in momentum of each body accumulated over the scattering process, 
$\Delta p_i^\mu = \int_{-\infty}^{\infty} \mathrm{d}\tau_i m_i \ddot{x}_i^\mu$.
In this work we restrict attention to the conservative sector of binary dynamics, where total momentum is conserved, $\Delta p_1 + \Delta p_2 =0$.
The scattering angle $\chi$ is then determined from the impulse via the relation
\begin{align}\label{def-chi}
2 \sin{\chi \over 2} = \frac{\sqrt{-\Delta p_1^2}}{p_{\infty}},
\end{align}
with $p_\infty$ the incoming center-of-mass momentum defined as
\begin{align}\label{}
p_{\infty}= {\mu\sqrt{\gamma^2-1} \over \Gamma}, 
~
\Gamma \equiv \frac{E}{m} = \sqrt{1+2 \nu(\gamma-1)},
\end{align}
where $\gamma = u_1\cdot u_2$ is the relativistic Lorentz factor characterizing the relative motion, $E$ and $m = m_1 + m_2$ are the  energy and total mass, $\mu = m_1m_2/m$ is the reduced mass and $\nu = \mu/m$ is the symmetric mass ratio.

\newpage
\vskip 5pt
\noindent
{\bf Impulse and scattering angle.}~Within the EFT framework, the impulse at high PM order is expressed as a linear combination of Feynman loop integrals.
Beyond the topologies already present in general relativity \cite{Kalin:2020fhe}, scalar-tensor theory introduces new topologies due to additional nonlinear couplings between the scalar and the worldline vertices, as listed in Figure \ref{3PM-topo-new}.
Notably, all of these additional diagrams can be viewed as sub-diagrams of certain graphs already encountered in the general relativity calculation of \cite{Kalin:2020fhe}. Consequently--and as expected--after performing integration-by-parts reduction using \texttt{FIRE6.5} \cite{Smirnov:2023yhb,Lee:2013mka}, the master integrals essential to this study fall entirely within the same set that arise in \cite{Kalin:2020fhe}.
Following the same analysis as in \cite{Kalin:2020fhe}, we obtain the following analytic results for the impulse to $\mathcal{O}(G^3)$:
\begin{align}
\Delta p_1^\mu = \sum_{n} {\nu m c_{n} \hat{b}^\mu  + d_n w^\mu \over J^n},
\quad
{1\over J} = {Gm \over |b|},
\end{align}
with
\begin{align}
\hat{b}^\mu = {b^\mu \over |b|},
~
w^\mu = 
{(\gamma m_1 + m_2)u_1^\mu - (\gamma m_2 + m_1)u_2^\mu  \over \sqrt{\gamma^2 - 1}},
\end{align}
where $b^\mu \equiv b_1^\mu {-} b_2^\mu$ is the impact parameter vector and $|b| = \sqrt{-b^2}$.
Here $d_1=0$ and the other coefficients up to $n=3$ are given by
\begin{widetext}
\vskip -18pt
\begin{align}\label{Dp-coefficients}
c_1 =& - \frac{2 (2 \gamma ^2 -1 +s_1 s_2)}{\sqrt{\gamma^2-1}},
\quad
d_2 
= -{\nu c_1^2 \over 2\sqrt{\gamma^2 - 1}}
\nonumber\\[0.5 em]
c_2 =  &   
- \frac{3\pi\, h_9}{8 \sqrt{\gamma ^2-1}}
+
{m_1 \over m}
\frac{ \pi   s_1 \left(s_1 s_{2}' +2 s_2 \left(s_1 s_2-2\right)\right)}{2 \sqrt{\gamma ^2-1} }
   +\frac{\pi m_1 s_1^2  \sqrt{\gamma ^2-1} }{4 m}
   - {3\pi\bar{\alpha} f_1 \over J^2}
   { m_1  \over m}
   \frac{ h_5 s_1+s_2}{2 \sqrt{\gamma ^2-1}  }
   + (1\leftrightarrow 2)
\nonumber\\[0.5 em]
d_3 = & 
-\frac{3 \pi\nu  h_{25}  }{4(\gamma ^2-1)^{3/2}}
+
   {\pi \nu  m_1 \over m}
\bigg[
\frac{   s_1 \left(2 s_1 \left(s_{2}' (h_1+s_1 s_2) + 2 s_2^2 (h_6+s_1s_2)\right) - h_{16} s_2\right)}{2 (\gamma^2-1)^{3/2}} 
+\frac{ s_1^2 \left(h_1+s_1 s_2\right)}{2 \sqrt{\gamma ^2-1}}
   \bigg]
   \nonumber
   \\
   &
   - {3 \pi \bar{\alpha} f_1 \over J^2}
   {\nu  m_1 \over m}
   \frac{ h_{23} s_1+s_2\left(s_1(h_5 s_1+s_2) + h_1\right)}{(\gamma^2-1)^{3/2}  }
   + (1\leftrightarrow 2)
\nonumber\\[0.5 em]
c_3^{(0)} =  &
- {m_1^2 \over m^2}
   \frac{6 h_{17}h_{18} + 2s_1 s_2 \left(s_1 s_2 (h_{21} s_1 s_2 - 3 h_{24}) + h_{27}\right)}{3 (\gamma ^2-1)^{5/2} }
   +
   {2  s_1^2 m_1^2 \over m^2}
   \bigg[
   \frac{ s_{2}' (h_8-h_6 s_1 s_2)}{(\gamma ^2-1)^{3/2}}
   +
  \frac{ s_1 (4s_2-s_{2}'')+4 h_1}{3 \sqrt{\gamma ^2-1}}
   \bigg]
   \nonumber
   \\
   &
+ {16 \bar{\alpha} \over J^2}
{m_1^2 \over  m^2}
\bigg[
\frac{  s_1 \left(f_1 \left(5s_{2}'+2 h_1 s_1^2\right)+f_2 \left(4 h_5 s_1+3s_2\right)\right)}{15 \sqrt{\gamma ^2-1} }
   -\frac{ f_1 \left(2 h_{29} s_1+s_2 \left(s_1\left(6 h_{13} s_1-5 h_{12}s_2\right)+h_{15}\right)\right)}{15 (\gamma^2-1)^{3/2} }
   \bigg]
   \nonumber
   \\
   &
   - {512 \bar{\alpha}^2 f_1^2 \over J^4} {m_1^2 \over m^2}
   \frac{ s_1 \left(h_7 s_1+3 s_2\right)+h_5}{35 \sqrt{\gamma ^2-1} }
   + (1\leftrightarrow 2)
\nonumber\\[0.5 em]
c_3^{(1)} =\,  &
2 \bigg[
   \frac{s_1 s_2 \left(s_{1}'  (s_{2}'+4s_2^2) + 4 s_1^2 s_2^2-2 s_1 s_2\right)+2h_{19}}{\gamma ^2-1} {\log x}
   - 8 s_1 s_2 \log x
   \nonumber\\
   &
   \quad+
   \frac{ \gamma s_2^2 s_{1}' (h_1+s_1 s_2)}{(\gamma^2-1)^{3/2}}
   +\frac{2 \gamma  s_2^2 (h_2+3 s_1 s_2)}{3 \sqrt{\gamma ^2-1}}
   - \frac{\gamma  \left(h_{32}-3 s_1 s_2 (s_1 s_2 (h_1s_1 s_2+h_{22}) + h_{20})\right)}{3 (\gamma^2-1)^{5/2}}
   \bigg]
   \nonumber\\
   &
+ {16 \bar{\alpha} f_1 \over  J^2}
\bigg[
\frac{ \gamma     s_2 \left(6h_4 s_{1}'+h_{14} s_1 s_2+h_{26}\right)}{3 (\gamma^2-1)^{3/2}}
   -\frac{2  s_2 \left(s_{1}'+2 h_{10} s_1 s_2-2 h_3h_{11}\right)}{(\gamma ^2-1)^2}\, \log x
   \bigg]
   \nonumber
   \\
   &
   + {64 \bar{\alpha}^2f_1^2 \over J^4}
   \bigg[
   \frac{3    \left(2 s_2 \left(h_{31} s_1+3 h_2 s_2\right)+h_{34}\right)}{(\gamma^2-1)^3} {\log x}
   +\frac{ \gamma  \left(2 s_2 \left(h_{28} s_1+h_{30} s_2\right)+h_{33}\right)}{(\gamma^2-1)^{5/2}}
   \bigg]
   + (1\leftrightarrow 2).
\end{align}
\end{widetext}
As anticipated, the result features the hallmark logarithm $\log x$ with $x= \gamma - \sqrt{\gamma^2 - 1}$.
For clarity, we have decomposed $c_3$ into two terms, $c_3 = c_3^{(0)} + \nu c_3^{(1)}$, which separately capture the behavior in the probe limit $\nu \to 0$.
Here the notation $(1\leftrightarrow 2)$ denotes the operation of exchanging the labels of the two particles, including their masses and sensitivity parameters.
We have also introduced the dimensionless parameter $\bar{\alpha} \equiv \alpha (Gm)^{-2}$ to characterize the Gauss-Bonnet coupling strength\footnote{Current constraints from gravitational-wave observations place an upper bound on $\sqrt{\alpha}$ of order $G M_\odot \simeq 1.5\,\mathrm{km}$, with $M_\odot$ the solar mass \cite{Lyu:2022gdr}.
For astrophysical-mass binary systems, this implies that $(Gm)^2$ captures the characteristic scale of $\alpha$, implying $\bar{\alpha} \sim 1$ for typical binaries.}.
The coefficients $h_i$ are polynomials in $\gamma$ up to degree 8, listed explicitly in Table \ref{table-h-coeficients}.

\vskip 4pt
Using the impulse presented above and the relation \eqref{def-chi}, we can directly extract the scattering angle.
The explicit form to $\mathcal{O}(G^3)$ is given in terms of coefficients $c_1$, $c_2$ and $c_3$ by
\begin{align}
{\chi \over \Gamma} =  
{2 \over J} & {|\lambda| \over \gamma^2 - 1}
- {1 \over J^2} {\lambda \over |\lambda|} \frac{c_2 }{\sqrt{\gamma^2 - 1}}
\\
+& {1 \over J^3} 
\bigg(
{4 \over 3}
\frac{\Gamma ^2 |\lambda|^3}{(\gamma^2 - 1)^3}
- {\lambda \over |\lambda|} \frac{c_3 }{\sqrt{ \gamma^2 - 1}}
\bigg)
+ \mathcal{O}(1/J^4),
\nonumber
\end{align}
where the second and third terms in the expansion depend on the sign of $\lambda = 2\gamma ^2 -1 +s_1 s_2$, determined by the magnitude and sign of the product $s_1 s_2$.

\vskip 4pt
While the general relativity contribution is already well-established, our work provides genuinely new corrections from the scalar-tensor and ESGB gravity (explicit expressions are also available in the ancillary file in machine-readable form).
Setting the sensitivity parameters to zero recovers the pure general‑relativity result.
The parameter $\alpha$ characterizes the contributions from the ESGB sector:~these enter the results starting at the next-to-leading order (NLO) through the interaction $\alpha\phi(\partial h)^2$, while the next-to-next-to-leading order (NNLO) receives also contributions proportional to $\alpha^2$.
As noted earlier, if the scale of $\alpha$ is identified with $(Gm)^2$, the Gauss-Bonnet couplings will effectively shift to higher PM orders.

\vskip 4pt
We have performed several non-trivial consistency checks to validate our results.
First, we verified that our results are independent of the choice of gauge-fixing term in \eqref{EFT}.
We independently carried out computations in many different gauge choices, including the de Donder gauge and a generalized gauge with arbitrary parameters, and found complete agreement among all cases.
Second, the impulse presented above satisfies the on-shell condition:~$(m_i u_i + \Delta p_i)^2 - m_i^2$ vanishes to $\mathcal{O}(G^3)$.
Finally, as a further nontrivial validation, we expanded the scattering angle in the post-Newtonian limit and confirmed that the scalar-tensor contribution reproduces the known results in~\cite{Jain:2023vlf}, which incorporate the 3PN-accurate effective-one-body coefficients computed in~\cite{Julie:2022qux}.

\begin{table}[t]
\begin{tabular}{|l|}
\hline
\parbox{55mm}{
\begin{align*}
&
\begin{aligned}
h_1 &= 2 \gamma^2-1\\
h_2 &= 2 \gamma^2+1\\
h_3 &= \gamma^2-5\\
h_4 &= \gamma^2-2\\
h_5 &= 3 \gamma^2-1\\
h_6 &= 2 \gamma^2-3\\
h_7 &= 4 \gamma^2-1\\
h_8 &= 4 \gamma^2-3\\
h_9 &= 5 \gamma^2-1\\
h_{10} &= 2 \gamma^2+3\\
h_{11} &= 4 \gamma^2+1\\
h_{12} &= 2 \gamma^2-5\\
h_{13} &= 9 \gamma^2-4
\end{aligned}
~~~~
\begin{aligned}
h_{34} &= 8 \gamma^6 (\gamma^2 {-} 1)+36 \gamma ^4+166 \gamma ^2+17\\
h_{33} &= 44\gamma ^6-94 \gamma ^4+356 \gamma ^2+351\\
h_{32} &= 20 \gamma ^6-90 \gamma ^4+120\gamma ^2-53\\
h_{31} &= 8 \gamma ^6+56 \gamma ^4+112 \gamma ^2+13\\
h_{30} &= 14 \gamma ^4-4 \gamma^6-16 \gamma ^2+33\\
h_{29} &= 58 \gamma ^4-49 \gamma ^2+6\\
h_{28} &= 46\gamma ^4+274 \gamma ^2+247\\
h_{27} &= 20 \gamma ^4-46 \gamma ^2+17\\
h_{26} &= 16\gamma ^4-22 \gamma ^2-243\\
h_{25} &= 10 \gamma ^4-7 \gamma ^2+1\\
h_{24} &= 8 \gamma^4-12 \gamma ^2+7\\
h_{23} &= 6 \gamma ^4-5 \gamma ^2+1\\
h_{22} &= 4 \gamma ^4-4 \gamma^2+3
\end{aligned}
\\
&
\begin{aligned}
h_{14} &= 28 \gamma^2-97\\
h_{15} &= 28 \gamma^2-13\\
h_{16} &= 31 \gamma^2-11\\
h_{17} &= 4 \gamma^3-4 \gamma -1\\
\end{aligned}
~~~~
\begin{aligned}
h_{21} &= 4 \gamma ^4-14 \gamma ^2+7\\
h_{20} &= -2 \gamma ^4+6 \gamma^2-1\\
h_{19} &= -4 \gamma ^4+12 \gamma ^2+3\\
h_{18} &= 4 \gamma ^3-4 \gamma +1
\end{aligned}
\end{align*}
}
\\
\hline
\end{tabular}
\vskip -5pt
\caption{Coefficient polynomials $h_i(\gamma)$.}
\label{table-h-coeficients}
\vskip -10pt
\end{table}

\vskip 5pt 
\noindent
{\bf Conclusions.}~In this work, we have initiated the analytic study of the conservative dynamics of compact binaries in massless scalar-tensor and Einstein-scalar-Gauss-Bonnet gravity in the post-Minkowskian approximation.
By combining the EFT approaches with modern multi-loop techniques, we derived gauge-invariant expressions for the momentum impulse and the scattering angle up to third order in the PM expansion.
These results provide new high-precision analytic ingredients for gravitational waveform modeling beyond general relativity, essential for precision tests of gravity and searches for new physics with current and future gravitational-wave observations.

\vskip 4pt
As anticipated, the EFT approach employed are directly applicable to theories beyond general relativity with the same robustness and flexibility as in the general relativity context. 
Therefore, extending the present calculations to higher perturbative orders is a natural and feasible next step. 
In particular, a complete computation at the fourth PM order appears well within reach, since the requisite master integrals are already available from earlier computations \cite{Dlapa:2023hsl,Dlapa:2021npj,Dlapa:2021vgp,Dlapa:2022lmu} and the relevant Feynman diagrams remain within manageable computational complexity.
We leave this for future work.

\vskip 7pt
\noindent
\emph{Added note}:~During the final preparation of this manuscript, the preprint \cite{Bernard:2026old} appeared on arXiv, reporting results in scalar-tensor theory that overlap with ours.
A direct comparison reveals a discrepancy between the two calculations.
Interestingly, we found that if one reverses the sign in the contribution from a specific diagram, their result is recovered.
However, we have thoroughly re-checked our analysis and are confident in the correctness of our results.

\vskip 6pt 
\noindent{\bf Acknowledgements.}~We thank Shuang-Yong Zhou for his collaboration during the initial stages of the project. This work was supported partially by the European Union’s Horizon Europe research and innovation programme under the Marie Sk\l{}odowska-Curie Action grant agreement No\,101146918 `{GWtheory}'.
This work was supported partially by the Start-up Research Fund of Southeast University No.\,RF1028624160, and China Postdoctoral Science Foundation No.\,2025M773334.

\newpage

\providecommand{\href}[2]{#2}\begingroup\raggedright\endgroup

\end{document}